\documentclass[aps]{revtex4}
\newcommand{\be}{\begin{equation}}
\newcommand{\ee}{\end{equation}}
\newcommand{\ba}{\begin{eqnarray}}
\newcommand{\ea}{\end{eqnarray}}
\newcommand{\bra}[1]{\left(#1\right)}

\usepackage{graphicx}% Include figure files

\begin{document}

\title{On Scaling Solutions with a Dissipative Fluid}
\author{J. Ib\'a\~nez}
\affiliation{Departamento de F\'{\i}sica Te\'orica,
         Universidad del Pa\'{\i}s Vasco,
Apdo. 644, 48080 Bilbao, Spain} \email{wtpibmej@lg.ehu.es}
\author{C. A. Clarkson}
\altaffiliation[Present address: ]{Relativity and Cosmology Group,
Department of Mathematics and Applied Mathematics, University of Cape Town,
Rondebosch 7701, Cape Town, South Africa}
\email{clarkson@maths.uct.ac.za}
\author{A. A. Coley}\email{aac@mathstat.dal.ca}
\affiliation{Department of Mathematics and Statistics, Dalhousie
University, Halifax, Nova Scotia  B3H 3J5, Canada}

\date{\today}

\begin{abstract}
We study the asymptotic behaviour of scaling solutions with a dissipative fluid
and we show that, contrary to recent claims, the existence of stable
accelerating attractor solution which solves the `energy' coincidence problem
depends crucially on the chosen equations of state for the thermodynamical
variables. We discuss two types of equations of state, one which contradicts
this claim, and one which supports it.
\end{abstract}
\maketitle

Current observational evidence suggests that the Universe is flat and
accelerating \cite{P}. Since the average energy-density of matter is relatively
low there should consequentially exist a dark energy smoothly distributed with
roughly the same magnitude as the matter energy. In general this requires
fine-tuning of the initial conditions and has become known as the `energy'
coincidence problem. The simplest explanation for this dark energy is a vacuum
energy density; however, due to the discrepancy between the ``observed'' value
of the cosmological constant and that predicted by particle physics \cite{W},
alternative mechanisms should be considered. One model that has recently
attracted attention explains the ``missing'' energy in terms of the existence
of a minimally coupled scalar field (``quintessence'' field) \cite{S}. By a
suitable choice of the scalar field interaction potential, the scalar field
evolves becoming the dominant contribution to the energy density today. These
type of solutions (matter scaling solutions) were first studied in \cite{R} and
their stability, for the exponential potential case, were analyzed in
\cite{ss}. However, since FLRW solutions with a perfect fluid and a scalar
field evolve towards a scaling solution that does not accelerate (unless matter
violates the strong energy condition), Chimento \emph{et al.}~\cite{Ch}
considered a dissipative fluid by means of the existence of a bulk dissipative
pressure. Bulk viscosity may be associated with particle production \cite{Z},
may be understood as frictional effects that appears in mixtures \cite{U} or
even to model other kinds of sources (e.g. the string dominated universe
described by Turok \cite{T} or a scalar field \cite{Ib}).

In \cite{Ch} the authors studied the effect of the bulk viscosity by using the
truncated Israel-Stewart theory \cite{I} and claimed that there exists a stable
attractor that has an accelerated expansion thereby solving the energy
coincidence problem. However, their analysis was not complete: the differential
equation for the function gamma defined in (10) of their article was not taken
into account and the equation of state for the bulk pressure was not stated. In
this paper we study the asymptotic behaviour of a FLRW model with a dissipative
fluid and a scalar field with exponential potential using two different
equations of state for the bulk pressure.

We will show that the claim stated in \cite{Ch}, that there exist stable
accelerating matter scaling solutions with non-zero bulk viscosity, cannot be
substantiated. We will show that the conclusions in~\cite{Ch} depend crucially
on the chosen equation of state for the coefficient of bulk viscosity, $\xi$,
and the relaxation time, $\tau$.

We assume a FLRW metric:
\begin{equation}
ds^2=-dt^2+a^2(t)\;d\Omega^2
\end{equation}
with a source given by a fluid with bulk viscosity and a scalar field with a
exponential potential:
\begin{equation}
T_{ab}=T^{\phi}_{ab}+T^m_{ab},
\end{equation}
where
\begin{equation}
T^\phi_{ab}=\phi_a\phi_b-g_{ab}\left(\frac{1}{2}\phi_c\phi^c+V(\phi)\right)
\end{equation}
and
\begin{equation}
T^m_{ab}=\rho u_a u_b+(p+\sigma)(u_a u_b+g_{ab}),
\end{equation}
where $\sigma$ is the viscous pressure, $p$ is the thermodynamic pressure and
$\rho$ is the energy density. Einstein's equations are then:
\begin{eqnarray}
\dot H & = & -H^2-\frac{1}{6}(\rho+3p+3\sigma+2\dot\phi^2-2V)\\
3H^2 & = & \rho+\frac{1}{2}\dot\phi^2+V-\frac{3K}{a^2}\qquad (K=\pm 1,0)\\
\dot\rho & = & -3H(\rho+p+\sigma)\\
\ddot\phi & = & -3H\dot\phi-\frac{dV}{d\phi},
\end{eqnarray}
where $H\equiv \dot a/a$.

The viscous pressure $\sigma$ is assumed to satisfy the truncated
Israel-Stewart equation \cite{I}:
\begin{equation}
\sigma+\tau\dot\sigma=-3\xi H,
\end{equation}
where $\xi$ is the coefficient of bulk viscosity and $\tau$ is the relaxation
time ($\xi>0, \tau>0$). Strictly speaking, eq. (9) is valid only when the fluid
is close to equilibrium; however, we assume its validity even when the fluid is
far from equilibrium. (A more rigourous and comprehensive analysis would
utilise the full transport equation of the Israel-Stewart theory.) We will
assume a linear barotropic equation of state $p=(\gamma-1)\rho$, where the
constant $\gamma$ satisfies $0<\gamma<2$  and two different relations for the
coefficient of bulk viscosity and relaxation times.

In the first case we consider the following relations:
\begin{equation}
  \qquad \xi=\alpha \rho^m, \qquad \tau=\frac{\xi}{\rho}=
 \alpha \rho^{m-1}
\end{equation}
which were introduced by Belinskii \emph{et al.} \cite{B}.

We now introduce  a new set of normalized variables:
\begin{eqnarray}
\Omega & = & \frac{\rho}{3H^2},\qquad \Sigma=\frac{\sigma}{H^2}, \qquad \Psi=
\frac{1}{\sqrt{6}}\frac{\dot\phi}{H},\nonumber\\
\Gamma & = & \frac{1}{3}\frac{V}{H^2},\qquad h=H^{1-2m},
\qquad m\ne \frac{1}{2},\label{vars}
\end{eqnarray}
and a new time $\tau$ defined by:
\begin{equation}
H(t)\;dt=d\tau.
\end{equation}

The Friedmann equation (6) now becomes:
\begin{equation}
1-\Omega^2-\Psi^2-\Gamma=-\frac{K}{H^2 a^2},\label{F}
\end{equation}
and  Einstein's equations are now:
\begin{eqnarray}
\Omega' & = & \Omega (-3\gamma+2x)-\Sigma\\
\Sigma' & = & -9\Omega+\Sigma\left[-\frac{1}{\alpha}(3\Omega)^{(1-m)}h+2x\right]\\
\Psi' & = & \Psi(x-3)-\frac{3k}{\sqrt{6}}\Gamma\\
\Gamma' & = & \Gamma (2x+k\sqrt{6}\Psi)\\
h' & = & -(1-2m)hx
\end{eqnarray}
where $'$ denotes derivative with respect to $\tau$ and
\begin{equation}
x\equiv 1+\frac{1}{2}(3\gamma-2)\Omega+\frac{1}{2}\Sigma+2\Psi^2-\Gamma.
\end{equation}

When $m=1/2$  the equation for $H$  decouples from the other equations and the
system becomes a four-dimensional system. We will not consider this particular
bifurcation situation just yet as it is a special case of the next equation of
state considered below.

Before studying the equilibrium points of the above system we present some
relevant expressions. The energy density of the scalar field is given by:
\begin{equation}
\rho_\phi=\frac{1}{2}\dot\phi^2+V=3H^2(\Psi^2+\Gamma)
\end{equation}
and
\begin{equation}
\Omega_\phi=\frac{\rho_\phi}{3H^2}=\Psi^2+\Gamma.
\end{equation}
The adiabatic index of the scalar field is:
\begin{equation}
\gamma_\phi\equiv\frac{\rho_\phi+p_\phi}{\rho_\phi}=\frac{\dot\phi^2}{\frac{1}{2}
\dot\phi^2+V}=\frac{2\Psi^2}{\Psi^2+\Gamma}.
\end{equation}
The deceleration parameter is given by:
\begin{equation}
q=x-1.
\end{equation}

We are looking for those equilibrium points that have non vanishing matter and
scalar field; i.e., $\Omega\ne 0$, $\Psi\ne 0$ and $\Gamma\ne 0$.  From
equation (17) one such point corresponds to:
\begin{equation}
x=-k\frac{\sqrt{6}}{2}\Psi.
\end{equation}
Equation (18) implies that $h=0$ and therefore $H\rightarrow0$ when $m<1/2$ and
$H\rightarrow\infty$ when $m>1/2$, as the point is approached. From  equation
(16) we obtain:
\begin{equation}
\Gamma=-\Psi^2-\frac{\sqrt{6}}{k}\Psi.
\end{equation}
From equation (15) we obtain:
\begin{equation}
0=-9\Omega+2x\Sigma,\qquad \Rightarrow \qquad \Omega=-k\frac{\sqrt{6}}{9}\Psi
\Sigma.
\end{equation}
Substituting this expression in equation (14) we get:
\begin{equation}
\Psi=\frac{\sqrt{6}}{4k}\left(-\gamma\pm\sqrt{\gamma^2+4}\right).
\end{equation}
Since $\Gamma>0$, equation (25) implies that $\Psi$ must be negative and
therefore only the solution with the minus sign is valid in the above
expression. Furthermore, it is easy to see that the condition  $\Psi<0$ implies
that $\gamma<3/2$. Finally, to obtain $\Sigma$ we substitute equations
(24)-(27) into equation (19) and we obtain:
\begin{equation}
\Sigma=-\frac{9}{\sqrt{6}k}\frac{1}{\Psi}\left(1+\frac{\sqrt{6}}{k}\Psi\right).
\end{equation}

So the equilibrium point is given by:
\begin{eqnarray}
\Omega & = & 1-\frac{3}{2k^2}\left(\gamma+\sqrt{\gamma^2+4}\right)\nonumber\\
\Sigma & = & \frac{6}{\gamma+\sqrt{\gamma^2+4}}\left[ 1-
\frac{3}{2k^2}\left(\gamma+\sqrt{\gamma^2+4}\right)\right]\nonumber\\
\Psi & = & -\frac{\sqrt{6}}{4k}\left(\gamma+\sqrt{\gamma^2+4}\right)\nonumber\\
\Gamma & = & \frac{3}{4k^2}\left[(2-\gamma)\left(\gamma+\sqrt{\gamma^2+4}\right)
-2\right]\nonumber\\
h & = & 0, \qquad \gamma<\frac{3}{2}.
\end{eqnarray}

The energy density of the scalar field is
\begin{equation}
\Omega_\phi=\Psi^2+\Gamma=\frac{3}{2k^2}\left(\gamma+\sqrt{\gamma^2+4}\right),
\qquad\Rightarrow\qquad \Omega+\Omega_\phi=1,
\end{equation}
and the adiabatic index is
\begin{equation}
\gamma_\phi=\frac{2\Psi^2}{\Psi^2+\Gamma}=\frac{1}{2}\left(\gamma+
\sqrt{\gamma^2+4}\right)\ne\gamma.
\end{equation}
The deceleration parameter in this case is given by:
\begin{equation}
q=\frac{3}{4}(\gamma+\sqrt{\gamma^2+4})-1>0.
\end{equation}
As in the matter scaling solutions, this equilibrium point represents a flat
universe in which the energy density of the scalar field scales with that of
the matter; however, in these solutions, the adiabatic indexes are different.
More importantly: this solution is not accelerating.

The stability of this equilibrium point is determined by the sign of the
eigenvalues of the matrix of the linearized system around the point. After a
lengthy calculation one can find that three of the eigenvalues are:
\begin{equation}
3\sqrt{\gamma^2+4},\qquad \frac{3}{2}(\gamma+\sqrt{\gamma^2+4})-2,\qquad
\frac{3}{2}(m-\frac{1}{2})(\gamma+\sqrt{\gamma^2+4}).
\end{equation}
The sign of the two first eigenvalues are positive whereas the sign of the
third depends on the value of $m$. Thus this equilibrium point is not stable.
The signs of the two other eigenvalues are difficult to determine.

It is important to note that this is the only equilibrium point with $\Omega\ne
0$, $\Psi\ne 0$ and $\Gamma\ne 0$. If we relax the last condition we obtain, in
addition to the former point, a set of equilibrium points corresponding to
$\Gamma=0$; i.e., a massless scalar field. From equation (16) we get $x=3$ and
from (18) $h=0$. From equations (14) and (15) we obtain:
\begin{equation}
\Sigma=\frac{3}{2}\Omega,\qquad \gamma=\frac{3}{2}.
\end{equation}
And finally from equation (19) we obtain:
\begin{equation}
\Omega+\Psi^2=1.
\end{equation}
In this case $\gamma_\phi=2$ and $\Omega_\phi=\Psi^2$,
($\Omega+\Omega_\phi=1$). Again this point is not stable since the eigenvalues
of the matrix of the linearised system are:
\begin{equation}
0, \qquad 6(m-\frac{1}{2}),\qquad 6+\sqrt{6} k \Psi,\qquad 4,\qquad 15.
\end{equation}

We shall now consider the second case by using the following equations of
state:
\begin{equation}\label{eos2}
\xi=\xi_0\rho^mH^{1-2m},~~~\tau=\tau_0\rho^{m-r}H^{2r-2m-1},
\end{equation}
as used in \cite{CvdH}.

The dynamical system is unchanged except for the evolution equation for the
viscous pressure, which now becomes, in terms of the new variables~(\ref{vars})
\begin{equation}\label{newsigma}
  \Sigma'=-a\Omega^r+\Sigma\left\{-b\Omega^{r-m}+2x\right\}
\end{equation}
where
\begin{equation}\label{}
  a=\frac{3^{1+r}\xi_0}{\tau_0},~~~{\mathrm{and}}~~~b=\frac{3^{r-m}}{\tau_0}
\end{equation}
replace our original positive constants $\xi_0$ and $\tau_0$.
 As in \cite{CvdH} we  take $r=1$, and we take
$m\leq r$ to ensure continuity of~(\ref{newsigma}). Note that as $H$ does not
appear in this equation the equation for $h$ decouples from the rest of the
dynamical system and we need only study the resulting four-dimensional system.
Note also that when $m=1/2$, we have a slight generalisation of the equation of
state used above.

There are two equilibrium points which satisfy the requirement $\Omega>0,
\Gamma\neq0,$ and $\Psi\neq0$. The first has
\begin{eqnarray}
\Sigma&=&-\Omega\bra{k^2\bra{\Omega-1}+3\gamma}\label{sigsol}\\
\Psi&=&\frac{1}{\sqrt6}k\bra{\Omega-1}\\
\Gamma&=&\bra{\Omega-1}\bra{\frac16k^2\bra{1-\Omega}-1},\label{gasol}
\end{eqnarray}
where $\Omega\in(0,1)$ must satisfy
\be
\bra{k^2\bra{\Omega-1}+3\gamma}\bra{b\Omega^{1-m}+k^2\bra{\Omega-1}}-a=0\label{omsols}
\ee
From~(\ref{gasol}) and $\Gamma>0$ we see that
\be
\Omega>1+\frac{6}{k^2}\label{omineq};
\ee
hence, from~(\ref{omsols}), we see that
\be
a<3\bra{\gamma+2}\bra{6+b\Omega^{1-m}}\label{aless}
\ee
must be satisfied for this point to exist.

The scalar field has density parameter and adiabatic index
\be
\Omega_\phi=1-\Omega,~~~\gamma_\phi=\frac13k^2\bra{1-\Omega}.
\ee
The deceleration parameter is given by
\be
q=\frac12k^2\bra{1-\Omega}-1=\frac32\gamma_\phi-1.
\ee

\begin{figure}[ht]
\includegraphics[width=10cm]{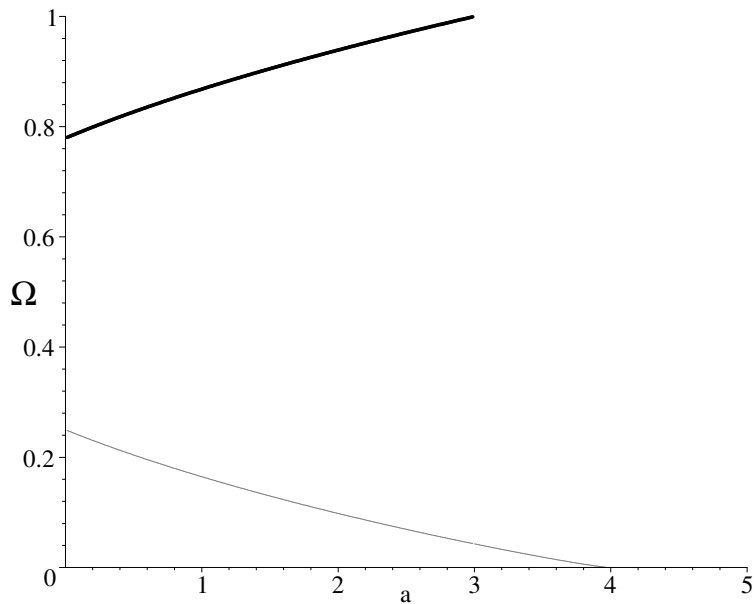}
\caption{\small A plot of $\Omega$ versus $a$ for the parameter
values $k=2,~m=0.5,~\gamma=1$ and $b=1$, obtained as the solutions
of~(\ref{omsols}). The thin grey curve (lower) represent solutions which are
unstable, while the thicker black curve represents the stable solution, and is
thus inflationary, by~(\ref{evs1}).
\label{b=1}}
\end{figure}

\begin{figure}[ht]
\includegraphics[width=10cm]{{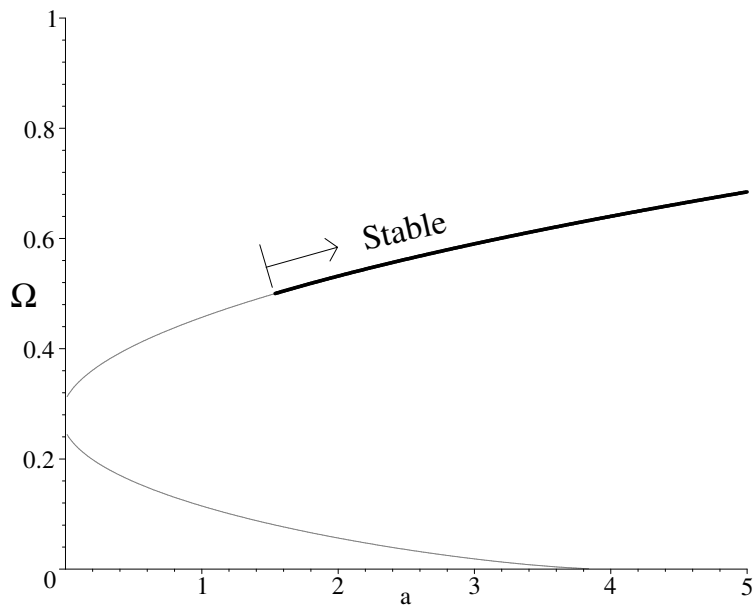}}
\caption{\small As in figure~(\ref{b=1}), but now with $b=5$.
Again, the thin grey curves represent solutions which are unstable, while the
thicker black curve represents the stable solution, and is thus inflationary.
This continues increasing with $a$ until $\Omega=1$ is reached when $a\sim14$
(cf. Figure~\ref{b=1_5_10}). \label{b=5}}
\end{figure}

\begin{figure}[ht]
\includegraphics[width=10cm]{{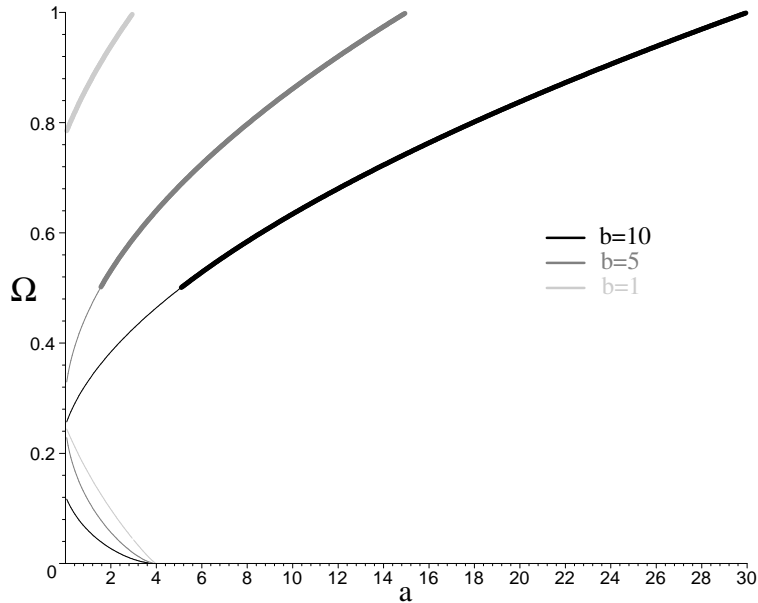}}
\caption{\small A plot of $\Omega$ versus $a$ for the parameter
values $k=2,~m=0.5,~\gamma=1$ and $b=1,~5,~10$, obtained as the solutions
of~(\ref{omsols}). This is an amalgamation of the previous two plots, with the
additional $b=10$ case represented in black, with the previous cases given in
shades of gray, as shown in the key. Again, the thick curves represent the
stable solution, and the thinner curves the unstable solutions. The case $b=10$
only becomes stable at high $a$. \label{b=1_5_10}}
\end{figure}

\begin{figure}[ht]
\includegraphics[width=10cm]{{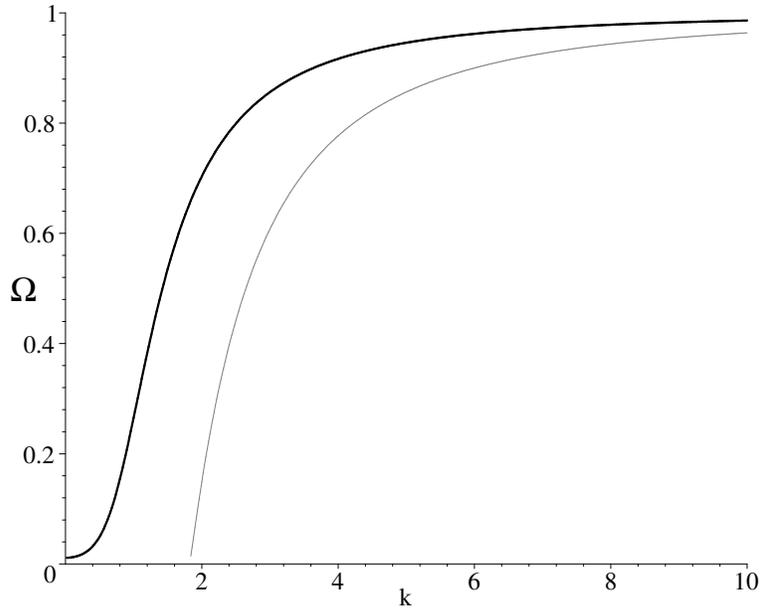}}
\caption{\small A plot of $\Omega$ versus $k$ for the parameter
values $a=1,~m=0.6,~\gamma=1$ and $b=1$, obtained as the solutions
of~(\ref{omsols}). The effect of increasing $k$ alters the value of $\Omega$,
but not the stability; the stable solution (the thick black curve) exists for
all $k$, while the unstable solution (thin, grey) becomes available for $k>2$.
\label{k}}
\end{figure}

This equilibrium point will be stable if all the real parts of the eigenvalues
of the Jacobian (of the rhs of the dynamical system) are negative. One of the
eigenvalues~$\lambda$ is easy to determine:
\be
\lambda=k^2(1-\Omega)-2=2q;\label{evs1}
\ee
thus if this point is not inflationary it will not be stable either. The
remaining eigenvalues may be determined but the resulting expressions are
extremely large. Therefore we will present the results for this point
graphically, in figures~(\ref{b=1}) --~(\ref{k}). As we only wish to
qualitatively show the different behaviour which may occur, we only show a few
representative cases, and show the variation of behaviour with the parameter
$a$. (Recall that $\tau_0=3^{1-m}/b$, and $\xi_0=a/3^{1+m}b$.) The effect of
increasing $b$ may be seen from the figures: for small $b$~(Fig. \ref{b=1}) the
upper curve (high density solution) is stable (as $a$ is increased) until
$\Omega=1$ is reached at $a\sim3$ [for $a$ larger than this the high density
solution no longer exists; cf. Eq.~(\ref{aless})]; similarly for $b=5$
(Fig.~\ref{b=5}), once the upper curve becomes stable (as $a$ is increased) at
$a\sim2$, it remains so until $\Omega=1$, which happens in this case at about
$a\sim14$. For the case $b=10$ (Fig.~\ref{b=1_5_10}), the higher density
solution becomes stable at $a\sim 6$, and $\Omega=1$ is reached at $a\sim30$.
The lower density solution appears never to be stable.

In the figures, we have arbitrarily set $\gamma=1$, $m=0.5$, and $k=2$. The
effect of increasing $\gamma$ or $m$ is to push the curves down to lower
$\Omega$ (the lower density solution eventually drops off the plots); this does
not change the conclusions. Lowering the value of $m$ allows the solution to
become stable for smaller $a$ (while pushing the curves up to higher $\Omega$);
e.g., in figure~(\ref{b=5}) with $m=0.1$, the upper curve becomes red (stable)
before $a=1$. The effect of changing $k$ is shown in figure~(\ref{k}).

The important point here is that stable solutions exist for a large non-trivial
region of the five-dimensional parameter space $\{a,b,m,k,\gamma\}$ of the
models.

%FIGURE 1

%FIGURE 2

%FIGURE 3

%FIGURE 4

\begin{figure}[ht]
\includegraphics[width=10cm]{{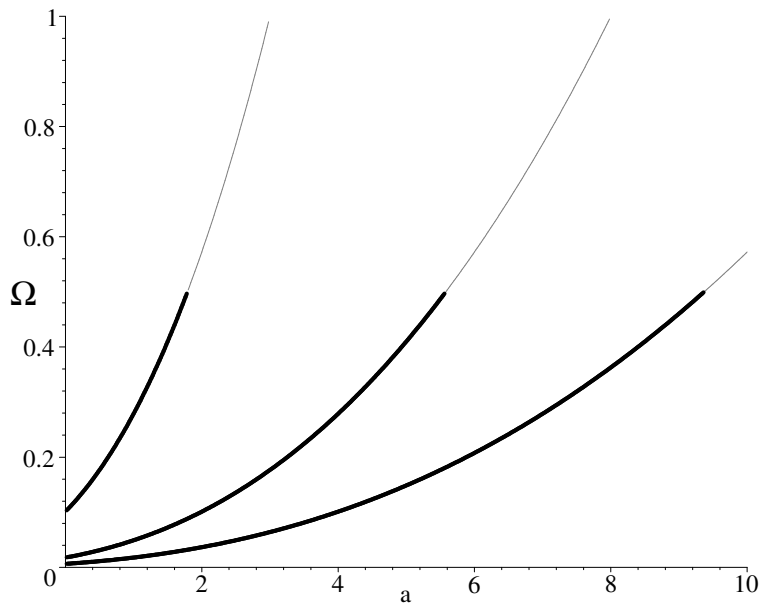}} \caption{\small A plot of
$\Omega$ versus $a$ for the parameter values $k=2,~m=0.6,~\gamma=1$ (so
$\Omega_\phi=0.5$); three values for $b$ are represented, $b=5,~10,~15$, with
$b=5$ being the upper curve, and $b=15$ the lower. The thin grey curves
represent solutions which are unstable, while the thicker black curves
represent the stable solutions. \label{3b}}
\end{figure}

The other equilibrium point of interest here is characterised by the following
(when $m\neq1$) :
\ba
\Omega&=&\bra{\frac{b\bra{3\gamma-2}}{a+2\bra{3\gamma-2}}}^\frac{1}{m-1},\\
\Sigma&=&-\Omega\bra{3\gamma-2},\\
\Psi&=&-\frac{\sqrt6}{3k},\\
\Gamma&=&\frac{4}{3k^2}.
\ea
For this point
\be
\Omega_\phi=\frac{2}{k^2},~~~\gamma_\phi=\frac23,~~~q=0.
\ee
Therefore this point is not inflationary. However, this point in general has
non-zero curvature. The stability of this point is difficult to determine
analytically, so again we present the results graphically in figure~(\ref{3b}).
Again we see that there exist stable solutions.

This graph~(\ref{3b}) suggests that, for these parameter values, all solutions
with $\Omega>0.5$ are unstable, a fact implied by further investigations
(although the limiting value of $\Omega$ depends on $k$: indeed for high $k$
there are no \emph{unstable} solutions). This then suggests that the stability
of the solution depends on the curvature: for $\Omega<\Omega_\phi=2/k^2~(=0.5$
in figure~(\ref{3b})) we have the stable solutions, and these have negative
curvature, by~(\ref{F}).

%FIGURE 5

For the previous two points we have demonstrated the existence of stable
solutions; however, we ignored the traditional thermodynamical assumption that
the magnitude of the viscosity must be less than the isotropic pressure (we
neglected this in the figures by setting $\gamma=1$). However, we note that
this does not affect the result. (So, e.g., in the latter equilibrium point
this requirement would set $\frac12<\gamma<\frac34$; stable solutions exist in
this case too.)

Finally, if we relax our restriction that $\Gamma\neq0$, we have another two
equilibrium points, given by (when $m\neq1$)
\ba
\Omega&=&\bra{\frac{3b(\gamma-2)}{a+18(\gamma-2)}}^{\frac{1}{m-1}},\label{om2}\\
\Sigma&=&3\Omega(2-\gamma),\\
\Psi^2&=&1-\Omega=\Omega_\phi.
\ea
For this point $q=2,$ and $\gamma_\phi=2$, so it is not inflationary. Three of
the eigenvalues are
\be
4,~~~3(2m-1),~~~6+\sqrt6 k\Psi;
\ee
hence this point is not stable.

We have shown that the equations of state play a crucial role in the final
evolution of non-perfect fluids with a scalar field. Therefore the claim that a
dissipative fluid minimally coupled with a scalar field can resolve the
coincidence problem, although suggestive, depends dramatically of the equations
of state and on the parameters associated with them.


\begin{thebibliography}{25}

\bibitem[1]{P} S.Perlmutter {\it et al.}, Astrophys. J. {\bf 517}, 565 (1999).

\bibitem[2]{W} S.Weinberg, Rev. Mod. Phys. {\bf 61}, 1 (1989).

\bibitem[3]{S} V.Sahni and A.Starobinsky, Int. J. Mod. Phys. {\bf D9}, 373
(2000); L.Parker and A.Raval, Phys. Rev. Lett. {\bf 86}, 749 (2001).

\bibitem[4]{R} B.Ratra and P.J.E.Peebles, Phys. Rev. D {\bf 37}, 3406 (1988).

\bibitem[5]{ss} E.J.Copeland, A.R.Liddle and D.Wands, Phys. Rev. D {\bf 57},
4686 (1998); A.P.Billyard, A.A.Coley and R.J. Van den Hoogen, Phys. Rev. D {\bf
58}, 123501 (1998).

\bibitem[6]{Ch} L.P.Chimento, A.S.Jakubi and D.Pav\'on, Phys. Rev. D {\bf 62},
063508 (2000).

\bibitem[7]{I} W.Israel, Ann. Phys.(N.Y.) {\bf 110}, 310 (1976); W.Israel and
J.Stewart, Ann. Phys.(N.Y.) {\bf 118}, 341 (1979).

\bibitem[8]{Z} Ya.B.Zeldovich, Sov. Phys. JETP Lett. {\bf 12}, 307 (1970).

\bibitem[9]{U} N.Udey and W.Israel, Mon. Not. R. Astron. Soc. {\bf 199}, 1137
(1982).

\bibitem[10]{T} N.Turok, Phys. Rev. Lett. {\bf 60}, 549 (1988).

\bibitem[11]{Ib} A.Di Prisco, L.Herrera and J.Ib\'a\~nez, Phys. Rev. D

\bibitem[12]{B} V.A.Belinskii, E.S.Nikomarov and I.M.Khalatnikov, Sov. Phy. JETP
{\bf 50}, 213 (1979).

\bibitem[13]{CvdH} A.A. Coley and R. van den Hoogen, Class. Quantum Grav. {\bf 12},
1977 (1995).

\end{thebibliography}
\end{document}